# Application of Quantitative Systems Pharmacology to guide the optimal dosing of COVID-19 vaccines.

Mario Giorgi, Rajat Desikan, Piet H. van der Graaf and Andrzej M. Kierzek

Certara QSP, Certara UK Limited, 1 Concourse Way, Sheffield, S1 2BJ, UK

**Optimal use and distribution of Covid-19 vaccines involves adjustments of dosing. Due to the rapidly-evolving pandemic, such adjustments often need to be introduced before full efficacy data are available. As demonstrated in other areas of drug development, quantitative systems pharmacology (QSP) is well placed to guide such extrapolation in a rational and timely manner. Here we propose for the first time how QSP can be applied real time in the context of COVID-19 vaccine development.**

The SARS-CoV-2 pandemic has catalysed a remarkable mobilisation in vaccine development. The virus genome was sequenced almost instantly after the first cases were identified and new vaccines entered clinical trials within a couple of months, followed by regulatory approval and rollout of national vaccination programs within a year. Most of these vaccines use platform modalities, in some cases approved for the first time, which will enable even more rapid updates following the discovery of new variants.

In the initial stages of Covid-19 vaccine development, there was little time for extensive optimisation of treatment regimen (i.e. dose amount, number of doses and dosing intervals). To date, most vaccines have progressed successfully from first-in-human studies to demonstration of efficacy in the wider population within months. However, often only after regulatory approval and roll-out in the real world does the critical importance of optimisation of dosing regimens become apparent, mainly due to the challenges of balancing limited supply with near-universal demand in the context of epidemiological and health-economical outcomes at local and international levels. For example, the United Kingdom Joint Committee on Vaccination and Immunisation (UK JCVI) recommended to extend the interval between the primary and booster doses from the originally-approved 3 or 4 weeks to 12 weeks (which at the time of the recommendation had not been tested), thus allowing single dose vaccination of twice the number of people in the first phase of the rollout[1]. Another potential example of a possible area for dose optimisation, both in terms of efficacy and supply-chain management, is the increased response reported for an arm of AZD1222 trial where half of the primary dose, followed by a booster dose was tested[2]. In addition, there is growing realisation that different vaccines may have to be combined, but it will not be possible to test all possible combinations in actual clinical trials in a timely manner[3].

We anticipate that the requirement for dose optimisation will also remain when the focus will shift to sustaining long-term Covid-19 vaccination programs in the light of emerging new strains for the virus. In addition, due to the wide-spread roll-out of vaccination programs and expected drop in Covid-19 incidence, it would become more difficult to run clinical trials in a timely manner. Recently, we described how quantitative systems pharmacology (QSP) is being used in immuno-oncology (IO) drug development to address similar challenges, i.e., decreasing access to sufficient number of clinical trial participants and the inability to explore all possible combination therapies and dosing regimens, in a timely manner[4]. We now propose that QSP can be used in a similar manner in Covid-19 vaccine development and present the first results demonstrating proof-of-principle.

QSP focusses on supporting drug development with mechanistic modelling and simulation of underlying biology. A typical QSP model consists of a pharmacokinetic module, describing absorption, distribution, and elimination of the drug, connected to a systems biology model quantitatively describing biology of the disease and mechanisms of drug action. The model (usually expressed as a set of Ordinary Differential Equations), is first parameterised with diverse literature and preclinical data usually available before the start of a drug development project. The model then extrapolates from these data and produces a first hypothesis about efficacious dosing regimens, often before clinical data are available. When a stage of clinical trial is completed, the model is validated, refined, and then applied for extrapolation, thus informing the next stage of the program. Recently, an increasing number of models have reached the maturity required to inform regulatory submission[5], with most applications in combination dose selection in immune-oncology[4]. In terms of regulatory acceptance, QSP follows the trajectory of physiologically-based pharmacokinetics (PBPK), where system-wide mechanistic models of physiology underlying pharmacokinetics are now routinely used *in lieu* of clinical trials on drug-drug interactions and recently also other fields. In an analogous manner, QSP models informed by a fast-expanding volume of pre-clinical and clinical data on Covid-19 immunology and vaccination are useful tools for optimisation of Covid-19 vaccine dosing regimens, especially in the context of increasing challenge of clinical subject recruitment and confounding factors.

Since 2017, the Immunogenicity QSP Consortium[6] has focussed on modelling formation of anti-drug antibodies (ADA), an unwanted immunological response to therapeutic proteins. We used the seminal model of Chen, Hickling and Vicini[7] as a starting point, expanded the physiological compartment structure, and created a platform model which has now been validated with ~20 clinical compounds. In the wake of the Sars-Cov-2 pandemic, we repurposed this model to Covid-19 vaccines. Since the basic biology of the humoral immune response is the same regardless of whether we simulate an unwanted ADA response to therapeutic proteins or desired immunogenicity to a vaccine antigen, we could quickly re-purpose and expand the model by developing a vaccine administration module (lipid nanoparticle (LNP) mRNA in first instance). This illustrates an important and at times underestimated feature of QSP – mechanistic platform models can be quickly applied across seemingly unrelated therapeutic areas, which share the same underlying fundamental biology. Likewise, pre-clinical and clinical data collected in seemingly unrelated projects can be integrated within a single QSP platform and contribute to confidence in its application.

Figure 1 illustrates possible application of the new QSP model to dose regimen selection in mRNA Covid-19 vaccines. Our example focusses on extrapolation of longitudinal antibody response from Phase I/II clinical trial data to dosing intervals, dose amounts and long-term vaccination, which have not yet been tested in actual clinical trials. We use the virtual population methodology[8] to generate ensembles of parameter sets (typically referred to as virtual patients), which fall within the range of observed patient variability in 120 days long anti-RBD IgG titer profiles collected by Widge et al.[9], for individual subjects treated with two 100 ug doses of mRNA-1273 vaccine administered with an interval of 28 days. Figure 1A shows both the calibration result and a virtual trial showing extrapolation beyond 120 days to examine response durability, as well as the predicted effect of a second 100 ug dose on antibody levels. Importantly, it is predicted that vaccination 11 months after the second dose still produces a burst of antibody synthesis, characteristic of a so-called booster effect, rather than a new, primary response. The calibrated model can then be used to examine different intervals between the first and second dose (Figure 1B). In agreement with the expectation of many immunology experts[1], and preliminary evidence from a subset of the AZD1222 Phase III trial[2], expansion of the dosing interval is predicted to increase antibody responses. Our model predicts a bell-shaped response, with an optimum between 7 and 8 weeks (Figure 1B). The 12 weeks interval proposed by UK JCVI is predicted to lead to lower immune response than this theoretical optimum, but the expected response is still higher than with the original 28 days interval. Importantly, in the model, the second dose given after 12 weeks still acts as a booster rather than a new primary dose. A potential downside of extending the time of the dosing interval is that antibodies may drop to low levels before the booster dose is administered. To explore this issue in a quantitative manner, we used convalescent plasma concentration as a reference (Figure 1A and 1B). While the question whether IgG level is a correlate of protection remains subject to debate and investigation, we note that the median convalescent IgG level is very close to the level observed in Covid-19 vaccine Phase III trials between day 10 and 14, the earliest timepoints where placebo and vaccine incidence curves separate[10]. The prolonged exposure of the virus to relatively low level of antibodies may also increase concerns related to the selection of vaccine-escaping mutant strains[1], although the modelling of *in vivo* virus mutation rates over 12 weeks would be needed before drawing conclusions.

Figures 1C and 1D illustrate an application of the QSP model to examine the effects of varying dose amounts in different age groups. While QSP is in principle applicable towards mechanistically modelling the aging immune system, we adopted a more phenomenological approach here and created two virtual populations calibrated by the clinical data of Walsh et al.[11], collected separately for younger (18-55) and older (65-85) adults. Using median convalescent concentration as a threshold, we calculated the percent of responding subjects at different time points (Figures 1C and 1D). Our results show that the antibody response is similar in age groups for 30 and 100 ug doses, consistent with observed high efficacy in older adults. However, lowering the dose to 10ug would have a larger negative impact in older adults (Figure 1D).

Another important application of large-scale mechanistic models is to generate virtual trials that enable the investigation of biomarkers, including (virtual) ones, which were not measured in the actual clinical trials. For example, Figure 2 shows the time profiles of memory B cells and memory CD4 T-cells in plasma simulated by the model calibrated with mRNA-1273 data (Figure 1A). These results can be used in two ways. First, the model can be subject to additional calibration by other biomarkers than IgG, thus increasing confidence in results. Second, model predictions can be used to guide selection of most informative biomarkers to be clinically investigated.

In summary, we believe that dose regimen optimisation will become increasingly important in ongoing and future development of Covid-19 vaccines. Is seems clear that the old "trial and error" vaccine development paradigm is inadequate to meet the worlds urgent needs. We therefore propose that, similar to other areas of drug development like IO[4], running virtual trials ahead of and in parallel with actual clinical trials using QSP models like the one presented here should become standard practise in vaccine development.

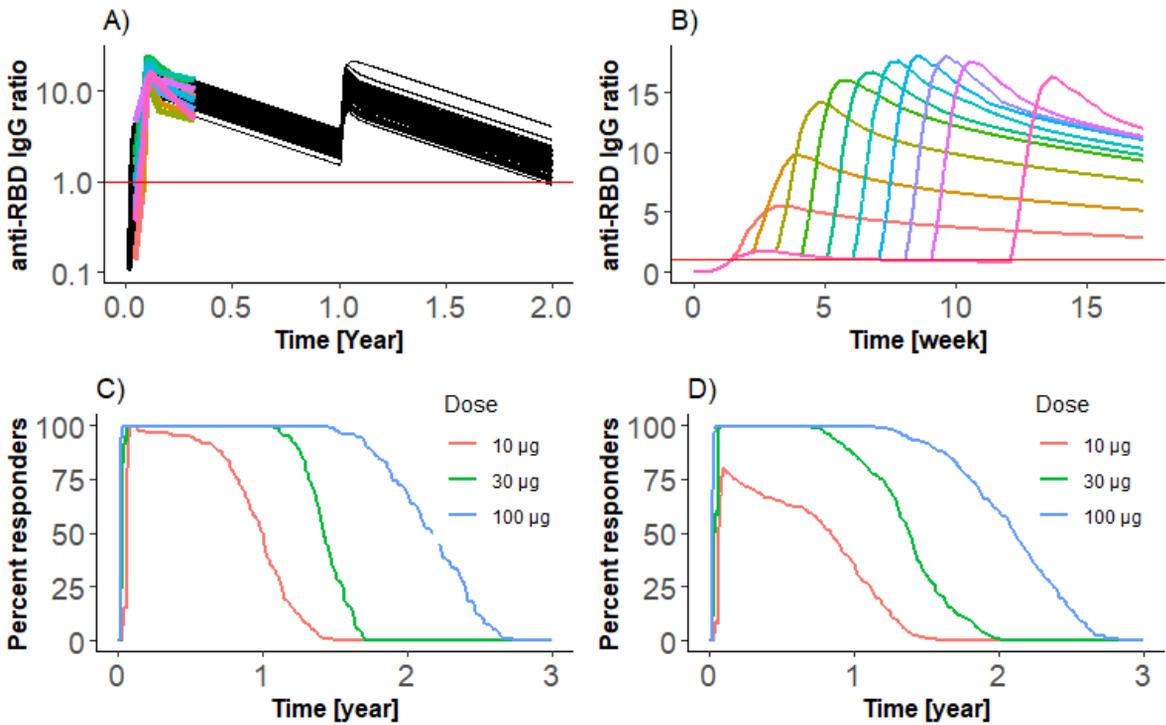

**Figure 1.** Example application of QSP vaccine model to extrapolate from Phase I/II data to different dosing regimens and long term vaccination. Plots A and B show ratio of anti-RBD IgG to the median of convalescent plasma concentrations, plotted by red horizontal line. A) Calibration with mRNA-1273 data and extrapolation to annual vaccination. Black lines show simulation results for 85 virtual patients. Coloured lines show clinical data available for first 120 days. A 100 ug dose was given at days 0, 28 and 365. B) Extrapolation to different intervals between primary and booster dose. The model calibrated for mRNA-1273 vaccine was used to predict antibody response for 100 ug dose administered at intervals of 1-9 and 12 weeks. We plot median IgG ratio of 85 virtual subjects. Administration of second dose leads to burst of antibody production, with maximum amount following bell-shaped curve. C,D) Extrapolation from Phase I/II data on BNT162b2 vaccine to different dose amount in younger (C) and older (D) adults. Two doses were given with 21 day interval and amounts of 10, 30, 100 ug. The 246 and 121 virtual patients were simulated in older and younger age groups. Plots show percent of virtual patients with anti-RBD amount above median convalescent plasma concentration at each time point. Response durability depends on the dose. The 10 ug dose results in substantially lower antibody response in older individuals.

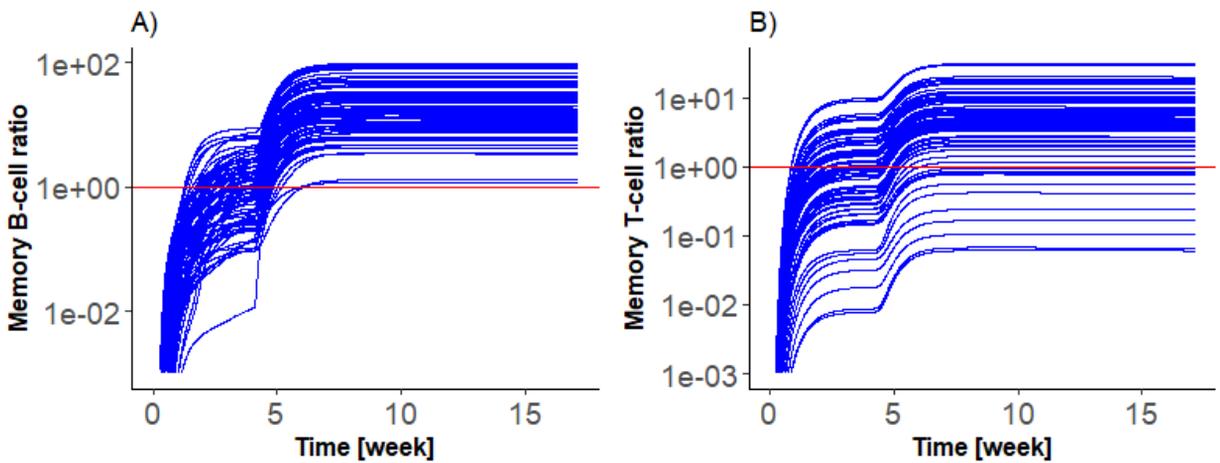

**Figure 2.** Example application of QSP model calibrated by Phase I/II data for investigation of biomarkers which were not observed in the clinic. The QSP model was calibrated by clinical data for anti-RBD IgG titer following administration of 100 ug mRNA-1273 vaccines to younger adults at day 0 and 28. Calibrated mechanistic model simulates not only antibodies, but also other biomarkers of interest. Here, we plot A) Memory B-cells and B) Memory CD4 T-cells in plasma compartment. Plots show ratios of the number of cells in plasma compartment, to the median number of cells at day 28, before booster dose was administered. Administration of booster dose increases both B and T cell memory. The model predicts considerable variability of individual responses, especially for T-cells.